# Roles in Software Development using Domain Specific Modelling Languages

Holger Krahn    Bernhard Rumpe    Steven Völkel
Institute for Software Systems Engineering
Technische Universität Braunschweig, Braunschweig, Germany
http://www.sse.cs.tu-bs.de

**Abstract**

Domain-specific modelling languages (DSMLs) successfully separate the conceptual and technical design of a software system by modelling requirements in the DSML and adding technical elements by appropriate generator technology. In this paper we describe the roles within an agile development process that allows us to implement a software system by using a combination of domain specific models and source code. We describe the setup of such a process using the MontiCore framework and demonstrate the advantages by describing how a group of developers with diverse individual skills can develop automotive HMI software.

## 1  Roles in a DSML-based development

Domain-specific modelling enables developers to separate previously connected development activities for a software system. Thus it allows them to concentrate on a single task at a time which leads to better results [4]. Furthermore, the development becomes more efficient, as parts of work can be reused from other projects more easily. In accordance to [4] we identify (in a simplified fashion) the following three activities during development:

- Domain specific modelling languages (DSMLs) are developed, reused or existing ones are enhanced to express the desired models of the problem domain.

- Code generators are implemented that transform models to an executable solution.

- The project specific knowledge or problem description is expressed in the DSMLs and the generators are used to map these models into a running solution.

These development activities are usually applied by different people according to their individual skills. By different code generators or even direct execution of the DSL instances the models are first class artefacts within the development. They can be used for different tasks like documentation, automated tests and rapid prototyping [16]. Therefore it is worthwhile to separate the activities mentioned above and assign them to specific roles:

- A *language developer* defines or enhances a domain specific modelling language (DSML) in accordance with the needs of the product developers.

- A *tool developer* writes code generators for the DSML which includes the generation of production and test code as well as the analysis of the content and its quality. In addition tool developers integrate newly-developed or reused language processing components and generators to form tools used within the project.

- A *library developer* develops software components or libraries and simplifies thereby the code generator because constant reusable software parts do not have to be generated. Therefore this role is closely connected to the *tool developer* but requires more domain knowledge. One aim of a library is to encapsulate detailed domain knowledge and provide a simplified interface that is sufficient for the needs of the code generation.

- The *product developers* use the provided tools for different activities within the project. Mainly, they specify a solution using their domain knowledge expressed in DSMLs to directly influence the resulting software.

The language developer not only defines the syntax of the modelling language respectively the newly added concepts in that language but also describes its meaning in terms of semantics, ensures that the new concepts are properly integrated in the existing language and provides a manual for their use. It is important that the semantics of a language is not only defined by describing how the generator handles it [9].

In conventional non-agile project settings both roles, language and tool developer, are not part of the project team. In cases where a commercial off-the-shelf tool is used, they are completely unavailable. However, the experiences we made so far indicate that it is recommended to integrate these tool based service activities into the project, leading to a more agile form of development.

The running system produced by the code generator allows the product developer to gain insights into the system's behaviour and thus gives the developers immediate feedback. Then the product developer might provide new feature requests and in turn the tool and language developers change the generator implementation or the DSML itself. In accordance to an agile development process, we argue that all developers should be able to easily adapt and immediately compile the resulting system after each change to judge the influence of the applied change easily. This requirement corresponds to the agile principles of *immediate feedback* and *continuous integration* [2]. This is only possible, if the language and tool developers are available within the project. Furthermore, in smaller projects the aforementioned roles might be taken by single person, thus the language, tool, library and product developer roles are unified.

The advantage of the development steps and roles is the strict separation of the description of a solution on a conceptual level and its technical realisation. This is an example of the well know principle *separation of concerns* [5] and permits the ability to independently evolve and possibly reuse all artefacts. The approach gains its benefit from the fact that technical solutions, stored in libraries and code generators change less often than the requirements for a certain application.

Our experience is based on the development of the MontiCore framework [8] which we use to develop DSMLs and tools which process these DSMLs. MontiCore itself is developed in an agile way, where the requirements of certain DSML descriptions (the input of MontiCore) often lead to changes in the generator itself and therefore evolve the MontiCore framework. MontiCore can be automatically rebuilt after any change in a MontiCore artefact and tests with different type of granularity ensure the quality of the result.

Our agile model-driven method uses a lot more concepts of other agile methods like *Extreme Programming* [2]: on-site customer, test-first, early feedback, etc. However, instead of a code-centric approach we concentrate on executable models that we use for production and test code generation. The main idea as described in [3] is to detect errors as soon as possible and to get early feedback from customers. Furthermore, the test cases we generate run in full automation, which makes the development process really agile. An on-site customer can act as a product developer that is not only able to develop the system in an appropriate way but can also define tests using the same notation [19]. All roles should develop their artefacts in a test-first manner regardless if they use or develop DSMLs or write source code. This makes an explicit test role unnecessary.

To explain such a development process in more detail we have developed a tool chain for a Human-Machine-Interface (HMI) in an automotive context. Two DSMLs are developed and used by different roles to produce an HMI based software.

The rest of the paper is structured as follows. Section 2 describes the MontiCore framework which enables an agile development of domain-specific modelling languages and the tool support for such a development process. Section 3 describes an illustrative example for the roles in the process where DSMLs are used for the development of an automotive sub-system. Section 4 relates our approach to other publications and Section 5 concludes this paper.

## 2  DSML-Framework MontiCore

As an intermediate step towards a fully model-based software development process we currently advocate a development process that uses code and models at the same time and, more important at the same level of abstraction. Several kinds of models and source code together express solutions in a problem-adequate way. This means, the developers do not round-trip engineer and switch views between code and models, but use models and handwritten code as descriptions of orthogonal parts. Developers do not look at or modify any form of generated code.

In [7] we have shown how to combine Statecharts and Java source code that exceeds the approach current CASE tools provide. The developer modifies only the handwritten source code and the Statecharts without considering the generated source code. This imposes syntactic constraints between source code and Statecharts, like called events in source code must be accepted by the Statechart and events have fitting parameters, which are directly displayed on basis of the Statechart and the handwritten source code. This makes a generation tool much more useable, compared to a situation, where errors have to be traced back from the generated source code to the model.

This approach is different from the OMG MDA[13] approach, because MDA describes the usage of models at different levels of abstraction and basically one-shot model transformations to transform each models from one level down to a less abstract level. The last transformation then results in source code that forms the software system. Manual changes in the generated models resp. source code are generally allowed and therefore, repeated generation becomes difficult if not impossible. Figure 1 sketches the generation/transformation process as seen by MDA that is also similar to a classic CASE tool approach compared to our process (b) where constraints are checked between models and source code. The handwritten source code is transferred to the generated source code and changed automatically where technical details are needed to interact correctly with the source code generated from models.

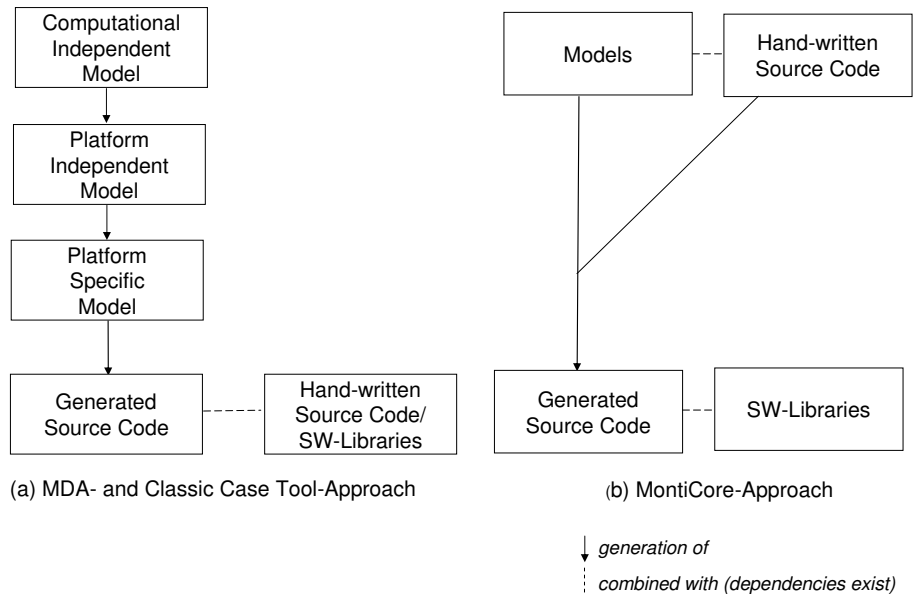

Figure 1: Comparison of MDA and the proposed approach

Executable UML [14] describes an approach where a well-defined subset of the UML is used to model a system. MontiCore completes this approach by additionally integrating a programming language as another kind of model and providing facilities to create new kinds of models.

MontiCore allows the language developer to define a modelling language by specifying the concrete syntax in form a context-free grammar. It uses this definition first for generating a lexer/parser pair with the parser generator Antlr [15]. In addition, it generates the internal representation (abstract syntax, metamodel) of the language as derivation from the grammar in form of Java classes. Through an extension mechanism within MontiCore grammars the standard derivation process can be flexibly adapted.

The language developer can express additional constraints and features that simplify the integration of the resulting products in the DSLTool framework of MontiCore. This framework provides standard solutions for tasks like file and error handling, execution order, tree traversal, template processing [12] or target code formatting. These techniques are a solid basis for the tool developer to define model transformations, code generation, analysis algorithms and syntactic checks based on the proper semantics and the intended use of the DSML. These solutions are offered to simplify the development of specific DSML tools within the agile development process.

Consequently, MontiCore can be seen as a generator on the one hand and as a language processing support environment on the other hand. The development of the MontiCore framework itself is a proof of concept for this approach, because the framework is implemented using a partial bootstrap process.

## 3  DSMLs for HMIs

This section demonstrates a practical example that uses the proposed MontiCore method for developing Human-Machine-Interfaces (HMIs) in cars. HMIs provide a user interface for the comfort functions of a car and are able to provide various feedback to the user.

Nowadays most car companies use their own HMIs with differences in look and feel, functions, and handling. Even cars of a single company have various configurations with different features. Taking the project setting of developing an HMI software for a certain car manufacturer with the agile MontiCore process, we have identified different activities in the development process and associate them with our identified roles.

The cooperation of the different artefacts can be found in Figure 2. The mentioned diagram types and languages are explained in the following.

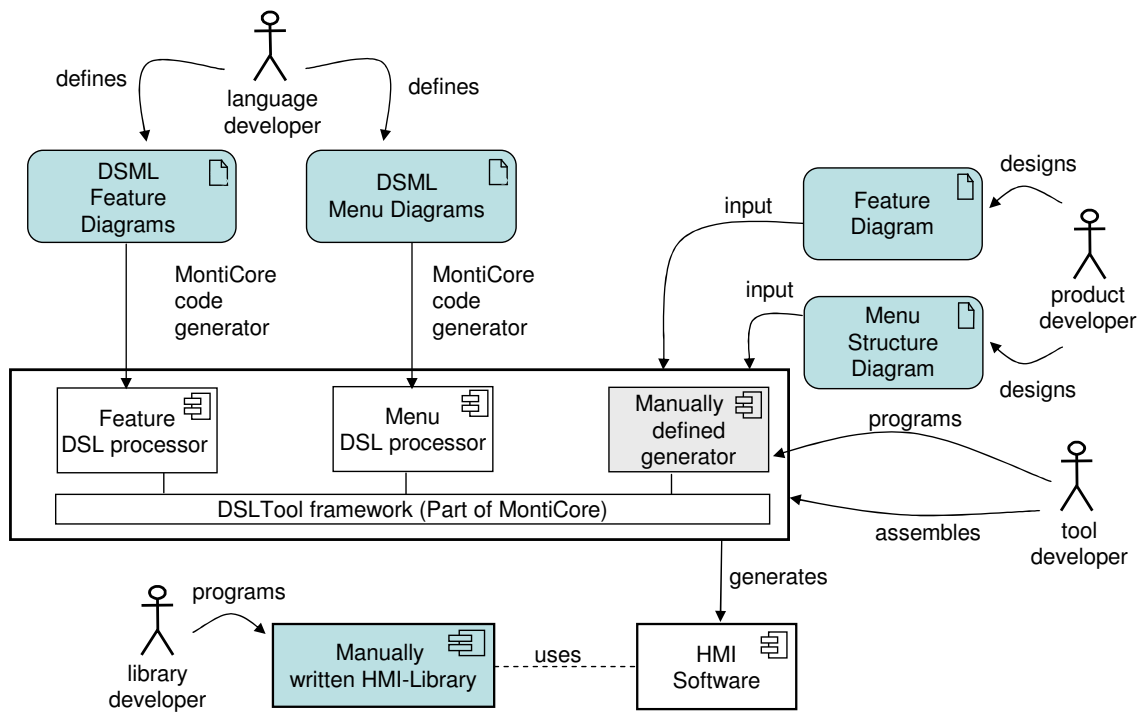

Figure 2: Generator structure for the HMI

After discussions with the product developers a *language developer* designs a DSML definition for *Menu Diagrams* that describe the menu structure of an HMI. This form of description is specific for HMIs in cars and uses concepts like menus, dialogs, status boxes and user inputs that correspond directly to the concept used by the manufacturer.

Another task for a *language developer* is to introduce *Feature diagrams* [4] to the project. These diagrams allow to model common and variable features and interdependencies between them. Figure 3 shows such a *feature diagram*. It is essentially a tree of features, that can either be mandatory or optional depending on the style of the edge: a black (mandatory) or a white (optional) circle. The edge decoration denotes alternative features. For the easier integration in the text-based tools in our proposed development process a textual notation for feature diagrams is used.

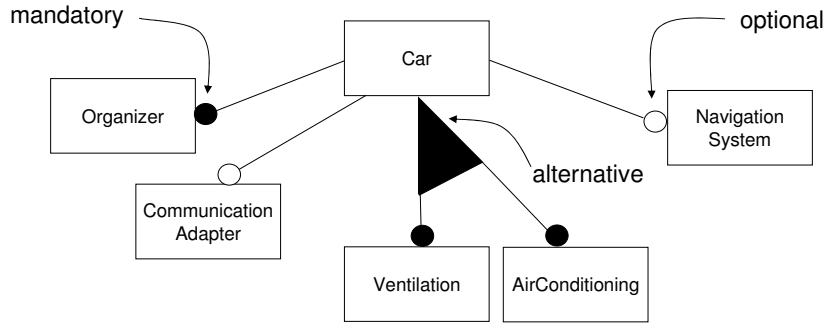

Figure 3: Feature Diagram

A *tool developer* builds a tool that comprises both languages. MontiCore is used to generate the language processing components and the DSLTool framework is configured to simplify the internal workflow and the input file handling of the tool. In addition a manually written code generator for HMI code is added to complete the tool.

The implementation of the generator is simplified by a *library developer* who develops an HMI-library that contains certain reusable code parts to program HMI software. The code generator simply configures the HMI library to form a specific HMI software.

The *feature modeller* describes feature sets which specify possible configurations of a type series and therefore is an instance of a *product developer*. An *HMI-developer* designs a menu structure for certain type series of cars. The HMI-developer therefore is another instance of a *product developer*. By using the developed tool and choosing a certain configuration for the car, he can directly generate the resulting software and simulate the result without further help of IT experts.

## 4 Related Work

Frequently *metamodelling* is used to create the abstract syntax of a modelling language. The Meta-Object Facilty [26] is the metamodelling technique standardised by the OMG where the metamodel is written as a simplified UML Class Diagram and OCL is used to define constraints on the abstract syntax. The MDA approach provides various ideas of integrating models into the development process which are primarily described as an input for one shot generations and therefore makes an agile process with continuous integration difficult. Due to the transformational nature of the approach the additional role *transformation definition engineer* is needed [11].

The Eclipse Modelling framework (EMF) [21] is another commonly used metamodelling framework. The meta-metamodel named Ecore can be used to create metamodels with the EMF framework itself, but also an import from a UML tool or textual notations like [10] and

[22] are possible options. Instances of the DSML can be created by a generic EMF editor. More sophisticated graphical editors can be either handwritten or created using the Graphical Modelling Framework (GMF) [24]. No strictly defined role based development process is proposed for the use with EMF.

The Generic Modeling Environment (GME) [23] is an integrated development environment for domain-specific modelling. The described MontiCore process could be adapted to be used with GME. A language developer would describe the abstract syntax of a language by a metamodel and define a graphical concrete syntax. GME is similar to MontiCore because a tool developer is supported by the environment to develop code generations or model interpretations. These artefacts can be reused inside GME to support product developers with an individually configured tool.

MetaCase's MetaEdit+ [25] uses a menu based editor to define metamodels. Models can be created through a graphical editor by drag and drop, inputs such as model names are made in input fields. MetaEdit+ uses its own *Report Definition Language* to navigate over a model instance and create code from it. The MetaEdit+ tool supports a variety of development processes and therefore does not go deep into process definition.

The *Domain-Specific Language Tools* [20] initiative from Microsoft also aims at the design of graphical DSMLs. The development is divided into three parts: definition of the metamodel, definition of (graphical) design, and definition of constraints. The meta-metamodel offers classes, value properties, and relations such as embedding (composition), reference (aggregation) and inheritance. Constraints are expressed in C#, code generation is supported by the *Template Transformation Toolkit* which allows an iterative access to DSL instances. Supported target languages for these templates are Visual Basic and C#. Sketches of an appropriate development process do exist e.g. in [6].

In [1] different roles for a model-driven development in general are presented. In comparison to our approach a more conventional software process is advocated with a separation in a meta and a project team. The paper mentions additional roles for testing and system analysis which are fulfilled by all developers in agile projects with activities like test-first design and constant feedback.

# 5 Conclusion

In this paper we have explained how an agile development process that uses code and models at the same level of abstraction can be used to efficiently develop a software system. We explained the different roles developers play in the realisation of a software when DSMLs are used to separate technological and application specific aspects. This technique also simplifies the integration of domain experts into a development team by giving them domain specific tools to express their knowledge without the need to go deeply into software issues.

The MontiCore framework strongly simplifies the development of DSMLs by providing an infrastructure the developer can rely on. This simplification is assisted by easy to use and quickly executed tools that enable a much more agile development process. Therefore, instead of a strict separation of tool and product developers, we are able to integrate those into the same project. In addition, we define new roles in a DSML-based project that will be carried out by developers respectively domain experts.

In the future we will further enhance the features of the MontiCore framework to be able to quickly develop more complex DSMLs. Furthermore, we will provide a number of predefined DSMLs that will serve as a basis for specific DSML definitions. Among others, we will develop a framework which supports UML/P [18, 17] as a special UML profile to model properties of a software for both, production and test code generation.